\documentclass{aa}
\usepackage{amsmath,graphicx,amssymb}
\usepackage{txfonts}
\usepackage{natbib}
\bibpunct{(}{)}{;}{a}{}{,}

\newcommand\hvezda{\object{$\theta$\,Aur}}
\newcommand\UBV{\textit{UBV}}
\newcommand\uv{\textit{uvby}}
\newcommand\oc{\textit{O-C}}
\newcommand{\zav}[1]{\left(#1\right)}
\newcommand{\hzav}[1]{\left[#1\right]}
\newcommand{\szav}[1]{\left\{#1\right\}}
\newcommand\intvidpo{\!\!\int\limits_{\begin{array}{c}\text{\scriptsize
visible}\\[-2mm]\text{\scriptsize surface}\end{array}}\!\!}

\newlength\staretab

\makeatletter
\def\sgn{\mathop{\operator@font sgn}\nolimits}
\makeatother

\begin{document}

\title{Visual and ultraviolet flux variability of the bright CP star $\theta$
Aur}

\author{J.~Krti\v{c}ka\inst{1} \and Z.~Mikul\'a\v sek\inst{1}
        \and T.~L\"uftinger\inst{2} \and M.~Jagelka\inst{1}}

\offprints{J.~Krti\v{c}ka,\\  \email{krticka@physics.muni.cz}}

\institute{Department of Theoretical Physics and Astrophysics,
           Masaryk University, Kotl\'a\v rsk\' a 2, CZ-611\,37 Brno, Czech Republic
            \and
            Institut f\"ur Astronomie, Universit\"at Wien,
            T\"urkenschanzstra\ss e 17, 1180 Wien, Austria
            }

\date{Received}

\abstract {Chemically peculiar stars of the upper part of the main sequence show
periodical variability in line intensities and continua, modulated by the
stellar rotation, which is attributed to the existence of chemical spots on the
surface of these stars. The flux variability is caused by the changing
redistribution rate of the radiative flux predominantly from the short-wavelength
part of the spectra to the long-wavelength part, which is a result of abundance
anomalies. Many details of this process are still unknown.}
{We study the nature of the multi-spectral variability of one of the brightest
chemically peculiar stars, $\theta$~Aur.}
{We predict the flux variability of $\theta$~Aur from the emerging
intensities calculated for individual surface elements of the star taking into
account horizontal variation of chemical composition. The surface chemical
composition was derived from Doppler abundance maps.}
{The simulated optical variability in the Str\"omgren photometric system and the
ultraviolet flux variability agree well with observations. The IUE flux
distribution is reproduced in great detail by our models in the near ultraviolet
region. A minor disagreement remains between the observed and predicted fluxes in
the far ultraviolet region. The resonance lines of magnesium and possibly also some
lines of silicon are relatively weak in the ultraviolet domain, which indicates
non-negligible vertical abundance gradients in the atmosphere. We also derive a
new period of the star, $P=3.618\,664(10)$\,d, from all available photometric
and magnetic measurements and show that the observed rotational
period is
constant over decades.}
{The ultraviolet and visual variability of $\theta$\,Aur is mostly caused by
silicon bound-free absorption and chromium and iron line absorption. Manganese
also contributes to the variability, but to a lesser extent. These
elements redistribute the flux mainly from the far-ultraviolet region to the
near-ultraviolet and optical regions in the surface abundance spots. The light
variability is modulated by the stellar rotation. The ultraviolet domain is
key for understanding the properties of chemically peculiar stars. It provides a
detailed test for surface abundance models and comprises many lines that originate
from states with a low excitation potential, which enable detecting
vertical abundance gradients.}


\titlerunning{Visual and ultraviolet flux variability of the bright CP star $\theta$\,Aur}
\authorrunning{J.~Krti\v{c}ka et al.}
\maketitle

\section{Introduction}

The atmospheres of main-sequence A and B stars are strongly affected by the
radiative force. The radiative force acts selectively on some heavier elements,
which leads to atomic diffusion and atomic segregation \citep{vaupreh,mpoprad}.
In normal AB stars the atomic diffusion is counteracted by the winds in hotter
stars and by turbulence in cooler stars \citep{mirivi}, consequently, their
abundance anomalies are only modest. The atmospheres of
chemically peculiar (CP) stars are more stable, however, and consequently show distinct
chemical composition remarkably different from the normal one \citep[see][for a
review]{rompreh}.

The magnetic field probably is the stabilizing force in a group of CP stars called
magnetic chemically peculiar stars. In these stars the magnetic field of the
order of hundreds of Gauss or higher dominates the whole atmosphere. Normal
A stars may possibly also have a surface magnetic field, but with an intensity lower by
two orders of magnitudes, as shown for Vega and Sirius
\citep{vegamag,sirmag}. Such weak fields do not significantly
affect the atmospheres. There seems to be a clear dichotomy between stars with strong and
weak magnetic fields \citep{dvojka}. Consequently, it seems that the so-called
normal AB stars and the peculiar ones are in many respects very similar apart
from the strength of their anomalies (the strength of the magnetic field and
the difference in chemical composition).

This view may be supported by the observation of the light variability
of the A star.
The light variability of CP stars is caused by the flux redistribution due to
bound-free \citep[ionization,][]{peter,lanko} and boud-bound
\citep[line,][]{vlci,ministr,molnar} transitions in surface abundance spots,
which is modulated with stellar rotation. It is tempting to attribute the
milli-magnitude variability observed in normal A stars \citep{balonek3} to the
same mechanism as in CP stars.

However, the test of this hypothesis is probably beyond the possibilities of current observational techniques. The situation in CP stars is much simpler. The theoretical light curves can be predicted using model atmospheres calculated
with actual abundances from abundance maps of individual stars. A comparison of
such theoretical light curves with observed ones supports the current picture of the light variability of
CP stars \citep{myhr7224,seuma,mycuvir}. This approach cannot
be used for rotationally variable normal A stars because the expected abundance
and line profile variations are prohibitively small for abundance mapping.

Consequently, studies of the light variability of  AB stars should concentrate on CP
stars with amplitudes of the order of hundredths of magnitudes. But our
knowledge is still far from complete even in this field. Most studies
concentrate on hotter CP stars, $\varepsilon$ UMa being the only example of a
cooler CP star studied so far \citep[based on maps of \citealt{leuma}]{seuma}.
Moreover, because we lack observational data in the UV, it is not always possible
to couple the study of optical light variability with that of
the UV spectral energy
distribution variability. However, this is crucial for CP stars, in which the
light variability originates in the UV, and the optical variations are just a
glimpse of it.

To fill this gap, we here provide a study of the optical and UV flux variability
of one of the brightest CP stars, \hvezda\ (HR 2095, HD 40312). This star belongs
to a group of well-studied CP stars. The Balmer line spectroscopy revealed the
influence of the Lorentz force on the stellar atmosphere of this star
\citep{sulymag}. The star is a frequent target of Doppler mapping
\citep[e.g.,][]{chochteta,ryze0,kuscoska,ryze}. We make use of the surface
abundance maps of various elements derived using Doppler images by
\citet{kuschnigthesis} and simulate the flux variability using model
atmospheres.

\section{Observations}
\label{pozor}

\hvezda\ has been recognized as an A0p Si star by \citet{cow}. The variability
in \textit{UBV} was established by \citet{winzer} and confirmed by two sets of \textit{uvby} measurements of \citet{adela} and \citet{aktax}. The star was also
observed by Hipparcos in the \textit{Hp} band \citep{esa97}. The measurements of the
effective magnetic field made by \citet{laborka,wadet}, and \citet{silvester} represent
further valuable phase information on the variability of \hvezda.

We supplemented these data by IUE spectra extracted from the INES database \citep[see
Table \ref{iuetab}]{ines} using the SPLAT package. We selected mostly
high-dispersion large-aperture spectra in the domains 1150--1900~\AA\ (SWP
camera) and 2000--3000~\AA\ (LWR camera). We used phase-dependent fluxes in our
analysis. Moreover, we calculated artificial magnitudes from the spectra by convolving them with a Gaussian filter with a dispersion of 25\,\AA\ (for SWP spectra)
and 100\,\AA\ (for LWR spectra).

\begin{table}[t]
\caption{List of the IUE observations of \hvezda}
\label{iuetab}
\centering
\begin{tabular}{ccccccc}
\hline
\multicolumn{3}{c}{SWP camera} & \multicolumn{3}{c}{LWR camera}\\
 Image & Julian date &  Phase & Image & Julian date &  Phase\\
        &   2,400,000+&&&2,400,000+\\
\hline
 21327 & 45630.77975 &  0.068 &  2111 & 45630.78658 &   0.070 \\
 21331 & 45631.04433 &  0.141 &  2135 & 45632.79204 &   0.624 \\
 21332 & 45631.07042 &  0.148 &  2139 & 45633.03122 &   0.690 \\
 21353 & 45632.79787 &  0.625 &  2150 & 45634.82937 &   0.187 \\
 21357 & 45633.03467 &  0.691 &  2154 & 45635.06043 &   0.251 \\
 21358 & 45633.06212 &  0.698 &  2166 & 45636.79184 &   0.729 \\
 21375 & 45634.79290 &  0.177 &  2184 & 45638.78399 &   0.280 \\
 21378 & 45635.06294 &  0.251 &  2186 & 45638.89943 &   0.312 \\
 21390 & 45636.78577 &  0.727 \\
 21412 & 45638.77651 &  0.278 \\
 21414 & 45638.90422 &  0.313 \\
\hline
\end{tabular}
\tablefoot{Phases were calculated according to Eq.~\eqref{naseef}.}
\end{table}

\section{Phenomenological modelling of the light and magnetic variability}\label{fenomodel}

The aim of the phenomenological modelling of the \hvezda\ light and magnetic field variability is to improve the rotational period of the star using all relevant sources of phase information, the few parametric descriptions of the light-curve shapes in the spectral domain from 1300 to 5500~\AA, and to determine the magnetic field geometry.

\subsection{Phenomenological model}

We found that each of the studied 19 light curves with effective wavelengths
from 1300 to 5500\,\AA\ is a smooth single wave curve, which can be approximated
by a harmonic function of second order. Nevertheless, the shapes of the light
curves in different bands are apparently different. We applied the weighted
advanced principle component analysis (APCA) to the parameters of the harmonic
fits, which allowed us to find hidden relationships among the individual light
curves \citep[for details see a brief introduction in][]{mikpca}. We concluded
that all light curves can well be expressed by a linear combination of only two
principle components and that the amplitudes of the third and higher principal
components are negligible.

\begin{figure}[tp]
\centering \resizebox{0.87\hsize}{!}{\includegraphics{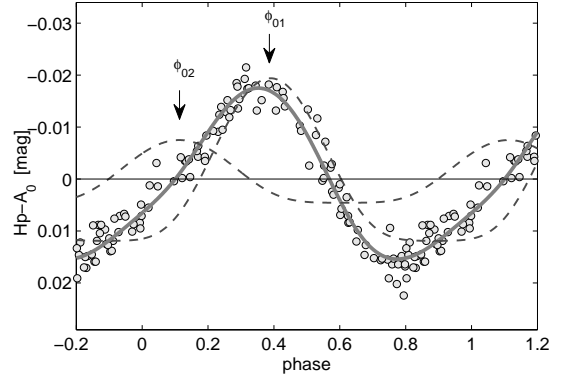}}
\caption{Light curve of \hvezda\ in \textit{Hp}
fitted by the model light curve (full line), which is the sum of two partial light curves
corresponding to individual spots (dashed lines) centred at $\varphi_{01}$ and
$\varphi_{02}$. Observations are denoted by circles.}
\label{Hp}
\end{figure}

This allowed us to build a relatively simple two-component phenomenological model
with a minimum of free parameters valid for all studied light curves, where each
of them is expressed by a linear combination of two principle normalized
symmetrical functions (see Fig.\,\ref{Hp}). The models of the light curves are
\begin{equation}
F(\vartheta,\lambda)=A_{0k}-2.5\,\log_{10} I_{\lambda},
\label{LC_model}
\end{equation}
where
\begin{equation}
I_{\lambda}= 1+\sum_{j=1}^2\,
A_{j\lambda}\szav{\exp\hzav{1-\cosh\zav{\frac{\varphi_j}{d_j}}}-2.288\,d_j}.
\end{equation}
The phase models are defined as
\begin{equation}
\displaystyle \vartheta=\frac{t-M_0}{P};\quad
\varphi_j=(\vartheta-\varphi_{0j})-\mathrm{round}(\vartheta-\varphi_{0j}), \quad
j=1,\,2,
\end{equation}
where $\vartheta$ is the so-called phase function \citep[sum of the common phase
$\varphi$ and the epoch $E$, for details see][]{brzda}, and $t$ is the time in
BJD. The origin of the phase function (zero phase time) $M_0$ was fixed at
$2\,450\,001.881$  \citep[the basic magnetic minimum according to][]{wadet}. $P$
is the period (assumed to be constant), $\varphi_{01}$ and $\varphi_{02}$ are
phases of centres of basic phase profiles (the phases of photocentres of
photometric spots), $d_1$ and $d_2$ are the parameters expressing the width of
the model spot profiles in phases, $A_{j\lambda}$ are the amplitudes of the
light variations for the $j$-th spot, $\varphi_j$ is the phase counted from the
centre of the $j$-th profile, and $A_{0k}$ is the mean magnitude of the $k$-th
set of observations from an individual author and an individual photometric colour.

The simplest phenomenological model of the phase curve of the longitudinal
magnetic field (corresponding to the oblique central dipole) is the cosinusoid
with the extrema at the phases when the magnetic poles are passing the stellar
meridian. Nevertheless, the recent observations of the longitudinal magnetic
field $B_l$ of \hvezda\ indicate a more complex phase behaviour \citep[see][and
also Fig.\,\ref{magnet}] {wadet,silvester}. That is why we approximated the
magnetic phase curve by the harmonic polynomial of the second order in the
following special form, which  allows us to estimate the extent of the distortion of
the basic purely cosine course:
\begin{align}\label{mag_model}
F(\vartheta)&=\overline{B}_{\mathrm l}+A_3\cos\zav{2\,\pi\,\vartheta_3}+A_4
\cos\zav{4\,\pi\,\vartheta_3}+\\
&+A_5\hzav{\sin(2\,\pi\,\vartheta_3)-\textstyle{\frac{1}{2}} \sin(4\,\pi\,\vartheta_3)},\quad \vartheta_3=\vartheta-\varphi_{03},\nonumber
\end{align}
where $\varphi_{03}$ is the phase of the magnetic curve minimum, $\overline{B}_{\mathrm l}$ is the mean longitudinal magnetic field, $A_3$ is the semi-amplitude of the cosine component of the magnetic variations. The semi-amplitudes $A_4   \text{ and}\,A_5$ quantify the significance of the possible symmetrical/antisymmetrical distortions of the observed phase curve.

\begin{table}[t]
\caption{List of the phase data sources used in the phenomenological modelling
of \hvezda\ variability. $E_{\mathrm m}$ is the mean epoch according to our
ephemeris (see Eq.\,\ref{naseef}). $N$ is the number of observations.}
\label{prehled}
\centering
\begin{tabular}{ccrrrl}
\hline
 Year &  &  $N$ & $E_{\mathrm m}$ & \oc\ \ \  &\ \ \ Source\\
\hline
 1972 & \UBV            &  45 &  -2372& -0.07(7) &  \tiny{\citet{winzer}} \\
 1977 & $B_{\mathrm l}$ &  19 &  -1931& 0.07(6) & \tiny{\citet{laborka}} \\
 1984 & IUE             &  106&  -1207& -0.05(10) & \tiny{this paper} \\
 1992 & \textit{Hp}     &  131&   -398& 0.03(3) &  \tiny{\citet{esa97}} \\
 1994 & \uv             &  188&   -210& -0.02(2) & \tiny{\citet{adela}} \\
 1998 & C &  11 &    209& -0.07(4)& \tiny{\citet{wadet}} \\
 2004 & \uv             &  276&    859& -0.01(2) &  \tiny{Adelman et al. (2005)}\\
  2008 &  $B_{\mathrm l}$      &  7&    1206& 0.00(2) &  \tiny{Silvester et al. (2012)}\\
\hline
\end{tabular}
\end{table}

The model describes all observed light curves obtained in 18 different
photometric colours and magnetic variations with a fairly high fidelity with only
67 free parameters. This phenomenological model was applied to
observational data representing 771 individual measurements. The measurements more or less evenly
cover the time interval 1971 -- 2008 (see Table~\ref{prehled}), and consist of
three sets of magnetic measurements, \UBV\ observations, two sets of \uv\
measurements, \textit{Hp} photometry, and 106 magnitudes in 11 ultraviolet wavelengths derived
from IUE spectra (see Sect.~\ref{pozor}).

We determined individual parameters $\boldsymbol a$ and their
uncertainties of the model described by Eqs.\,\eqref{LC_model} and
\eqref{mag_model} by deriving for each measurement its model prediction
$F(t_i,\boldsymbol{a})$ using the standard weighted least-square method
applied to the set of measured values (magnitude, magnetic field) $y_i$, with the estimated uncertainty $\sigma_i$,
\begin{align}\label{LSM}
\Delta y_i&=y_i-F(t_i,\boldsymbol{a});\quad \chi^2=\sum^n_{i=1}\zav{\frac{\Delta y_i}{\sigma_i}}^2;\nonumber\\
\boldsymbol{\nabla} \chi^2&=\boldsymbol{0}; \quad \Rightarrow\quad
\sum^n_{i=1}\frac{\Delta y_i}{\sigma_i^2}\
\boldsymbol{\nabla}\!_{\boldsymbol{a}}F(t_i,\boldsymbol{a}) =\boldsymbol{0}.
\end{align}
For details see also \citet{mikvar}.

\subsection{New ephemeris and \oc\ diagram}

One of the principle results of our phenomenological modelling was the improved   period of the observed stellar variations: $P=3\fd618\,664(10)$. The time of the zero phase and the zero epoch, $M_0=2\,450\,001.881,$ were adopted from the ephemeris published in \citet{wadet}. The times of the zero phase $\mathit{\Theta(E)}$ can be approximated by the relation
\begin{equation}\label{naseef}
\mathit{\Theta}(E)=2\,450\,001.881+3\fd618\,664(10)\,\times\, E,
\end{equation}
where $E$ is
the integer, also called the epoch. 
The period agrees well with other determinations \citep{wadet,aktax}.
The phase shift between this ephemeris and that used by \citet{kuschnigthesis}
in the time of their spectral observations is $\Delta\varphi=-0.268$.

We also tested a possible period variability. We assumed that the phase function
is approximated by the relation
\begin{equation}
\vartheta=\frac{t-M_0}{P_0}-\frac{\dot{P}}{2}\,\zav{\frac{t-M_0}{P_0}}^2;
\ \ \mathit{\Theta}=M_0+P_0\,E+\frac{\dot{P}\,P_0\,E^2}{2},
\end{equation}
where $P_0$ is the instant period at $M_0$ and $\dot{P}$ is the time
derivative of the period. We found that $\dot{P}=3(5)\times10^{-9}$, so we can conclude that there is zero quadratic term in the \hvezda\ ephemeris and the period is constant.

\begin{figure}[tp]
\centering \resizebox{0.915\hsize}{!}{\includegraphics{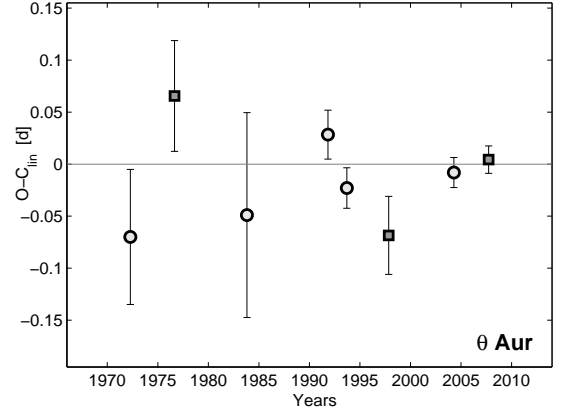}}
\caption{\oc\ diagram of \hvezda\ indicates that its period is nearly constant. Circles indicate \oc s derived from photometry, squares correspond to \oc s found from magnetic measurements.}
\label{oc}
\end{figure}

A more complex behaviour of the period can be revealed by means of the \oc\
diagram inspection. The phase function can be approximated by the relation
\begin{equation}
\displaystyle \vartheta(t,r)=\frac{t-M_0-(\oc)_r}{P_0},
\end{equation}
where $M_0$ and $P_0$ are fixed, $P_0=3\fd618\,664$, and $(\oc)_r$ are the mean
\oc\ values of the subsets $r$ of the observational data \citep[for details see in][]{mikoc}.
The \oc\ diagram in Fig.~\ref{oc} (constructed using observations
in Table~\ref{prehled}) does not indicate any period variations during the past 33 years.

\subsection{Light curves}

\begin{figure}[tp]
\centering \resizebox{\hsize}{!}{\includegraphics{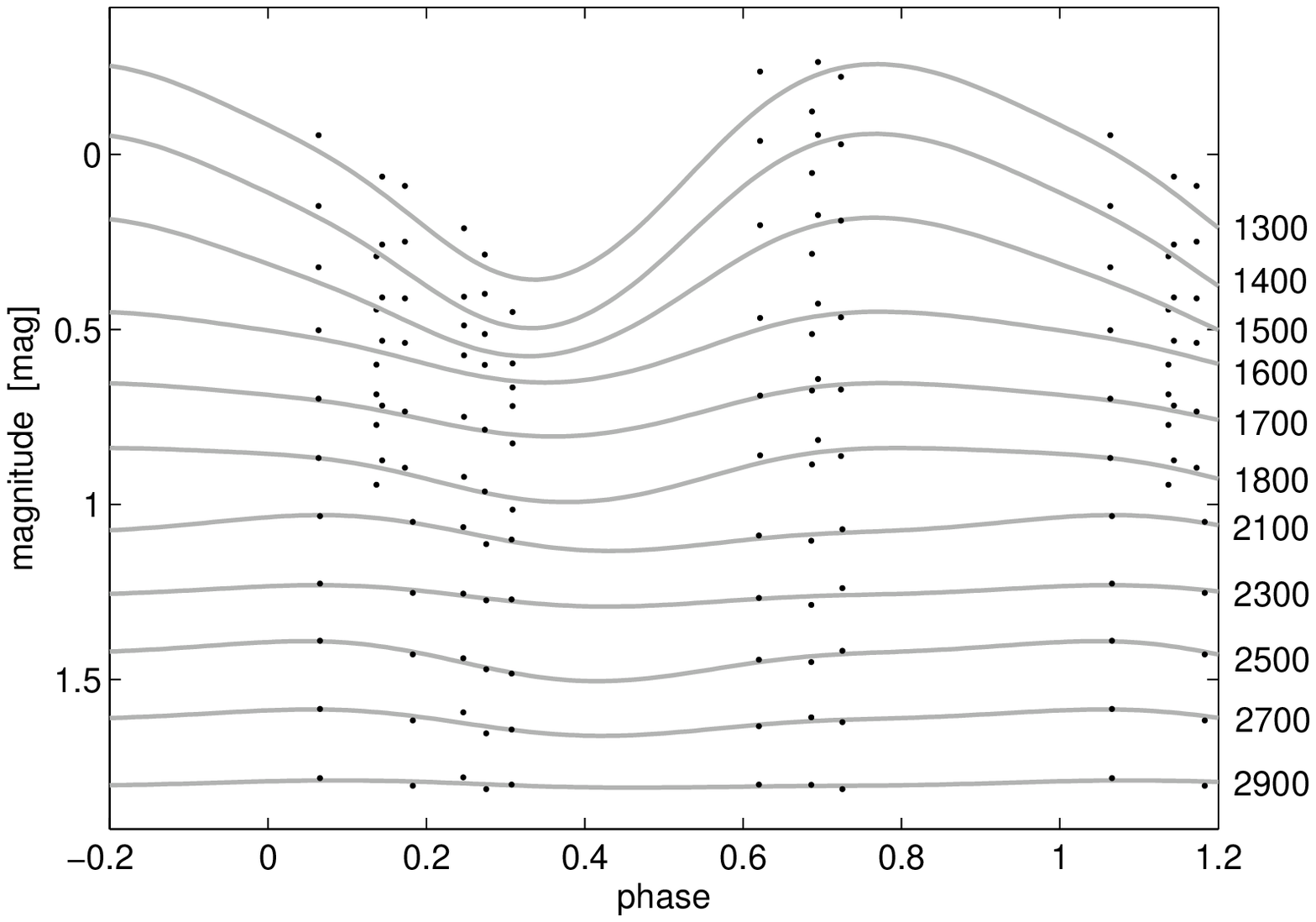}}
\centering \resizebox{\hsize}{!}{\includegraphics{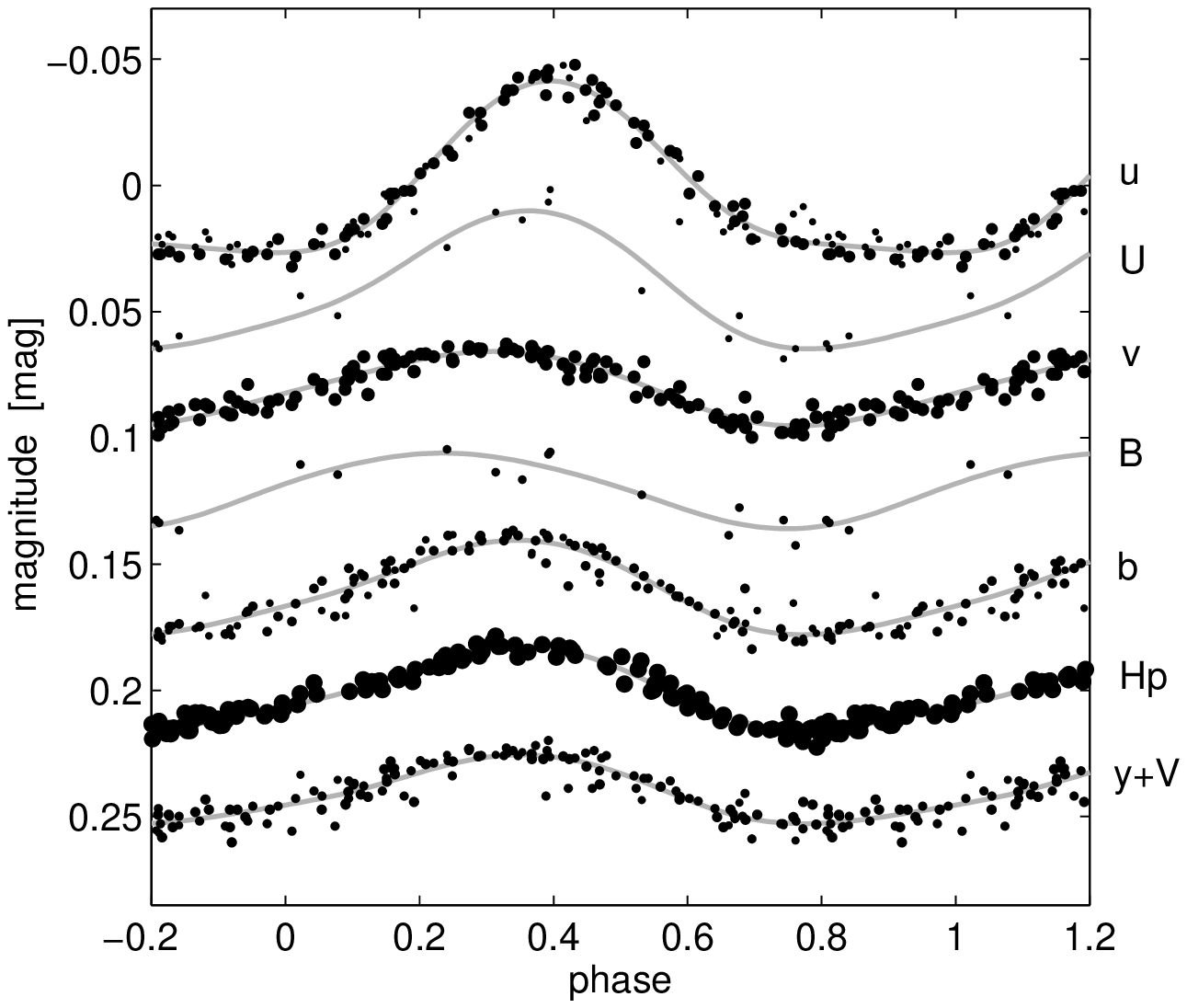}}
\caption{Observed light curves of \hvezda\ in UV ({\em upper panel}) and
in standard photometry ({\em lower panel}). Symbols on the right side denote the
effective wavelengths in \AA\ or a filter. Point areas in the lower panel
correspond to the weight of measurements. Grey lines are the results of phenomenological modelling.}
\label{krivkyvelke}
\end{figure}

\begin{figure}[tp]
\centering \resizebox{0.9\hsize}{!}{\includegraphics{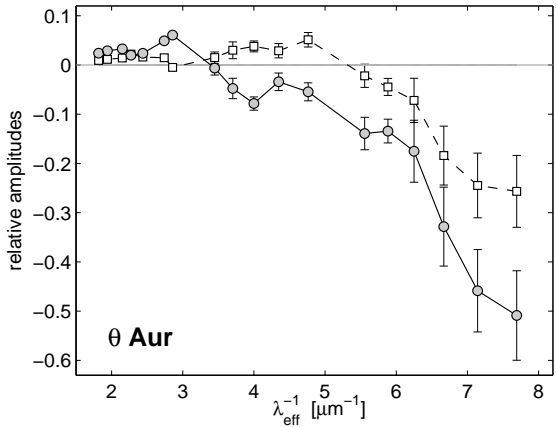}}
\caption{Amplitudes of two photometric spots centred at $\varphi_{02}=0.11$
(squares) and $\varphi_{01}=0.39$  (full dots) at various wavelengths.}
\label{amplvelke}
\end{figure}

We present the light curves of \hvezda\ in the region 1300 -- 5500\,\AA, which\ covers a substantial
part of the stellar flux, in Fig.\,\ref{krivkyvelke}. The largest
amplitudes are shown by the light curves at the short-wavelength end of the spectral region. The phenomenological modelling of the light curve set
by the linear combination of two basic symmetric phase profiles (see
Eq.\,\ref{LC_model}) was successful, the model is able to plausibly fit the
shapes of all observed light curves. The dominant feature of the spectrum is
centred at the phase $\varphi_{01}=0.389(4)$. Our test showed that the width
$d_2$ of the second profile centred at $\varphi_{02}=0.112(10)$ is nearly
identical to the width of the first profile. Consequently, we assumed that the
widths of all phase structures are the same, $d_1=d_2=d=0.165(3)$.

The observed behaviour of light curves can be interpreted by the presence of two
dominant photometric spots on the surface of the rotating star whose
photo-centres pass the stellar median at phases 0.389 and 0.110. The contrasts
of the spots with respect to their surroundings and the contrast of the spots
themselves determine the shape of the light curve in a particular photometric
colour. The principle spot is bright only for wavelengths longer than 3400\,\AA,\
otherwise it is dark. The second spot is bright for wavelengths longer than
1850\,\AA. For shorter wavelengths, both spots are dark, and together they create
one extensive shady feature.

\subsection{Geometry of the magnetic field and its relation to photometric spots}

\begin{figure}[tp]
\centering \resizebox{1\hsize}{!}{\includegraphics{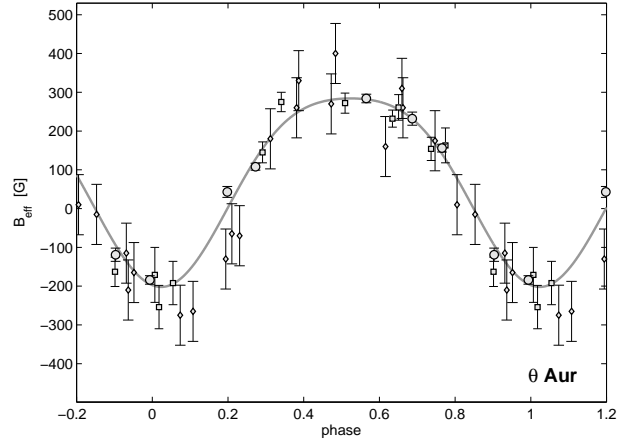}}
\caption{{Variations of the longitudinal component of the magnetic field of
\hvezda. Squares denote the observation of \citet{wadet}, diamonds the observation of
\citet{laborka}, and circles the observation of \citet{silvester}. The grey line is the symmetric model fit.}}
\label{magnet}
\end{figure}

The magnetic field variations are described by the model Eq.~\eqref{mag_model}.
The phase of the sharp magnetic minimum is $\varphi_{03}=0.024(4)$. The mean
value of the longitudinal component of the magnetic induction of \hvezda\ is
positive ($\overline{B}_{\mathrm l}=85(4)$\,G), which means that the positive
pole of the magnetic dipole is in the northern hemisphere of the star. The
derived semi-amplitude of magnetic variations is $A_3=-243(7)\,\text{G,}$ and the
semi-amplitude of the symmetrical component $A_4=-44(5)\,\text{G}$ is nonzero,
contrary to the semi-amplitude of the antisymmetric term $A_5=-8(10)$\,G. This
means that the phase curve of the magnetic variations is definitely distorted, but
symmetric around the phase of the magnetic minimum (see Fig.\,\ref{magnet}).
This form of the phase curve can be explained by assuming that
the magnetic dipole is slightly
shifted from the centre in the direction of the negative magnetic pole.

The phases of the centres of assumed photometric spots neither coincide with the
phases of magnetic field extrema nor with phases when the magnetic equator
passes the centre of the visible stellar disc. Moreover, the spots with the highest element
abundances mapped by \citet{kuschnigthesis} appear on the visible
disc around the phases $0.3-0.4$, again showing a mismatch with the surface
magnetic field. Thus we conclude that there is no obvious connection between the
photometric structures on the stellar surface and the stellar magnetic field
geometry.

\section{Calculating the flux variability}

\subsection{Model atmospheres and synthetic spectra}

The stellar parameters of \hvezda\ together with the range of surface abundances
of individual elements taken from \citet{kuschnigthesis} are given in
Table~\ref{hvezda}. The effective temperature and surface gravity were derived
from fitting the H$\delta$ lines. The projected rotational velocity was
derived from spectroscopy. The range of surface abundances was extracted from
surface abundance maps,  which were derived by Doppler imaging. In our study the
abundances are defined relative to hydrogen, that is,
$\varepsilon_\text{el}=\log\zav{N_\text{el}/N_\text{H}}$.

We used the code ATLAS12 for the model atmosphere calculations
\citep{kurat,casat}. For each model atmosphere we calculated the synthetic
spectrum using the code SYNSPEC with atomic data from \citet{bstar2006}. The
adopted SYNSPEC line list was enlarged by the data for chromium and iron taken
from Kurucz (2009)\footnote{http://kurucz.harvard.edu}. We computed
angle-dependent emerging intensities for $20$ equidistantly spaced values of
$\mu=\cos\theta$, where $\theta$ is the angle between the normal to the surface
and the line of sight.

In our models we assumed fixed effective temperature and surface gravity (see
Table~\ref{hvezda}) and adopted a generic value of the microturbulent velocity
$v_\text{turb}=2\,\text{km}\,\text{s}^{-1}$. The abundance of Mg, Si, Cr, Mn,
and Fe was selected according to the surface abundance maps (see below). We
assumed solar abundance of the other elements \citep{asp09}.

The model atmospheres and the angle-dependent intensities
$I(\lambda,\theta,\varepsilon_\text{Mg},\varepsilon_\text{Si},\varepsilon_\text{Cr},\varepsilon_\text{Mn},\varepsilon_\text{Fe})$
were calculated for a five-parametric grid of magnesium, silicon, chromium,
manganese, and iron abundances (given in Table~\ref{esit}). For silicon and iron
this grid fully covers the range of abundances found on the surface
of \hvezda\ 
(Table~\ref{hvezda}). For magnesium, chromium, and manganese the lowest
abundances were omitted from the grid because these elements with such low
abundance do not influence the emerging fluxes. This helped us to significantly
reduce the required computer time. Our test showed that the highest abundances
of helium, titanium, and strontium on the surface of \hvezda\ (see
Table~\ref{hvezda}) are so low that they do not influence the emerging flux.
Consequently, these elements were excluded from the grid of calculated
abundances in Table~\ref{esit}, and we only assumed a surface mean abundance of
these elements.

\begin{table}[t]
\caption{\hvezda\ parameters derived by \citet{kuschnigthesis}.}
\label{hvezda}
\centering
\begin{tabular}{lc}
\hline
Effective temperature ${{T}_\mathrm{eff}}$ & ${10\,500}$\,K \\
Surface gravity ${\log g}$ (cgs) & ${3.5}$ \\
Inclination ${i}$ & ${60^\circ}$ \\
Rotational velocity projection $v_\text{rot} \sin i$ & $55\,\text{km}\,\text{s}^{-1}$\\
Stellar radius\tablefootmark{1} $R_*$& $4.34(5)\,R_\odot$ \\
Helium abundance&$-2.8<\varepsilon_\text{He}<-1.9$ \\
Magnesium abundance& $-7.5<\varepsilon_\text{Mg}<-3.2$ \\
Silicon abundance& $-3.6<\varepsilon_\text{Si}<-2.8$ \\
Titanium abundance& $-7.9<\varepsilon_\text{Ti}<-6.9$ \\
Chromium abundance& $-6.0<\varepsilon_\text{Cr}<-3.4$ \\
Manganese abundance& $-6.2<\varepsilon_\text{Mn}<-4.7$ \\
Iron abundance&$-4.2<\varepsilon_\text{Fe}<-3.2$  \\
Strontium abundance&$-9.1<\varepsilon_\text{Sr}<-7.7$  \\
\hline
\end{tabular}
\tablefoot{
\tablefoottext{1}{Derived in Sect.~\ref{uv}.}}
\end{table}

\begin{table}[t]
\caption{Individual abundances $\varepsilon_\text{Mg}$, $\varepsilon_\text{Si}$,
$\varepsilon_\text{Cr}$, $\varepsilon_\text{Mn}$, and $\varepsilon_\text{Fe}$ of
the model grid}
\label{esit}
\centering
\begin{tabular}{lrrrrrrr}
\hline
Mg& $-4.1$ & $-3.6$ & $-3.1$ \\
Si& $-3.7$ & $-3.2$ & $-2.7$ \\
Cr& $-5.9$ & $-5.4$ & $-4.9$ & $-4.4$ & $-3.9$ & $-3.4$\\
Mn& $-5.7$ & $-5.2$ & $-4.7$ \\
Fe& $-4.2$ & $-3.7$ & $-3.2$ \\
\hline
\end{tabular}
\end{table}

\subsection{Phase-dependent flux distribution}
\label{vypocet}

\begin{table}[t]
\caption{Coefficients of analytical fits of the photonic passbands
of the Str\"omgren photometric system filters $uvby$ Eqs.~\eqref{besrov1} and
\eqref{besrov2}.}
\label{bestab}
\centering
\begin{tabular}{lrrrrrrrr}
\hline
& $a_1$ & $a_2$ &$a_3$ &$a_4$ &$a_5$ &$a_6$ & $d_c$ & $\sigma_c$\\
\hline
$u$ & 21.3 & -14.3 & 3590 & 8.85 & 448 & -1430 & 3475 & $10^3$\\
$v$ & 94.4 & -652 & 2000 & 55.2 & 305 & 0 & 4100 & $10^3$\\
$b$ & 76.7 & -268 & 0 &111 &-896 & 2790  &4675 & $10^3$\\
$y$ & 24.8 & 465 & 0 &47.1 &578 & 0 &5475 & $10^3$\\
\hline
\end{tabular}
\end{table}

The derived grid of spectra (Table~\ref{esit}) was used to calculate the radiative flux in a
bandpass $c$ at the distance $D$ from the star \citep{mihalas}
\begin{equation}
\label{vyptok}
f_c=\zav{\frac{R_*}{D}}^2\intvidpo I_c(\theta,\Omega)\cos\theta\,\text{d}\Omega.
\end{equation}
The intensity $I_c(\theta,\Omega)$ at angle
$\theta$ with respect to the normal at the surface was obtained at each surface point with spherical coordinates $\Omega$ by interpolating between the
intensities
$I_c(\theta,\varepsilon_\text{Mg},\varepsilon_\text{Si},\varepsilon_\text{Cr},\varepsilon_\text{Mn},\varepsilon_\text{Fe})$
calculated from the grid as
\begin{multline}
\label{barint}
I_c(\theta,\varepsilon_\text{Mg},\varepsilon_\text{Si},\varepsilon_\text{Cr},\varepsilon_\text{Mn},\varepsilon_\text{Fe})=\\
\int_0^{\infty}\Phi_c(\lambda) \,
I(\lambda,\theta,\varepsilon_\text{Mg},\varepsilon_\text{Si},\varepsilon_\text{Cr},\varepsilon_\text{Mn},\varepsilon_\text{Fe})\, \text{d}\lambda.
\end{multline}
The response function $\Phi_c(\lambda)$ of a given bandpass $c$ is a Gauss
function for the UV magnitudes (the same as for IUE fluxes, see
Sect.~\ref{pozor}). For the Str\"omgren photometric system we use analytical
fits to photonic passbands from \citet{specfce},
\begin{equation}
\label{besrov1}
\Phi_c(\lambda)=\left\{
\begin{array}{c}
\exp\{-x\hzav{a_{1c}+x(a_{2c}+a_{3c}x)}\},\quad x<d_c,\\
\exp\{-x\hzav{a_{4c}+x(a_{5c}+a_{6c}x)}\},\quad x>d_c,
\end{array}\right.
\end{equation}
where
\begin{equation}
\label{besrov2}
x=\frac{(\lambda-d_c)^2}{\sigma_c^2},
\end{equation}
where $\lambda$ is the wavelength in \AA.
The coefficients of the fits are given in Table~\ref{bestab}.

The magnitude difference between the flux $f_c$ at a given phase and the
reference flux $f_c^\text{ref}$ in bandpass $c$ is defined as
\begin{equation}
\label{velik}
\Delta m_{c}=-2.5\,\log\,\zav{\frac{{f_c}}{f_c^\mathrm{ref}}}.
\end{equation}
The reference flux is obtained under the
condition that the mean magnitude difference over the rotational period is zero.

\section{Influence of heavier elements on atmospheric structure}
\label{kaptoky}

\begin{figure}[tp]
\centering \resizebox{\hsize}{!}{\includegraphics{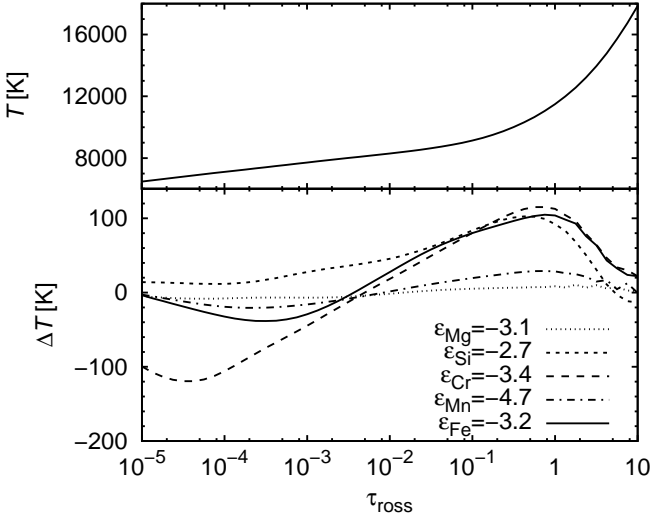}}
\caption{{\em Upper plot:}
Temperature dependence on the Rosseland
optical depth $\tau_\text{ross}$ in the reference model atmosphere with
$\varepsilon_\text{Mg}=-4.1$, $\varepsilon_\text{Si}=-3.7$,
$\varepsilon_\text{Cr}=-4.4$, $\varepsilon_\text{Mn}=-5.7$, and
$\varepsilon_\text{Fe}=-4.2$. {\em Lower plot}:
Difference between
temperature in the model atmospheres with
abundance of the selected element enhanced by a factor of $10$
and the temperature in the reference model atmosphere.}
\label{tep}
\end{figure}

\begin{figure*}[tp]
\centering \resizebox{0.75\hsize}{!}{\includegraphics{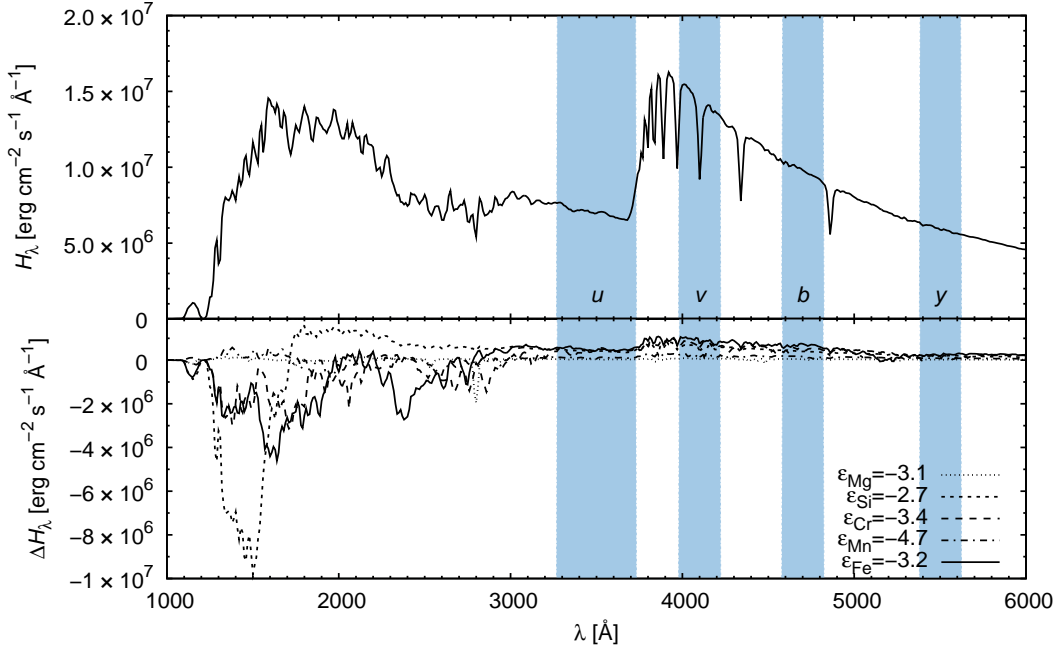}}
\caption{{\em Upper plot:}
Emerging flux from a reference model
atmosphere ($\varepsilon_\text{Mg}=-4.1$, $\varepsilon_\text{Si}=-3.7$,
$\varepsilon_\text{Cr}=-4.4$, $\varepsilon_\text{Mn}=-5.7$, and
$\varepsilon_\text{Fe}=-4.2$). {\em Lower
plot}:
Difference between the emerging flux from the model atmospheres with
higher abundance of the selected element by a factor of $10$
and the flux from a reference model. All
fluxes were smoothed by a Gaussian filter with a dispersion of $10\,\AA$ to show
the changes in continuum. The passbands of the $uvby$ photometric system are
also denoted.}
\label{prvtoky}
\end{figure*}

Heavier elements significantly affect the structure of stellar atmospheres.
Although their abundance is much smaller than that of hydrogen even in CP stars,
their contribution to opacity is crucial. Bound-free transition of silicon and
numerous line transitions of iron group elements (in our case, chromium and iron)
are able to absorb the stellar radiation, which leads to heating of continuum-forming regions ($\tau_\text{ross}\approx0.1-1$). This is shown in
Fig.~\ref{tep}, where we plot the temperature difference between the models
with enhanced abundance of a selected element by a factor of $10$
and a reference one.

The heavier elements also lead to a redistribution of the emerging
flux typically from the far UV to the near UV region and to the optical
domain. This is shown in Fig.~\ref{prvtoky}, where we compare the
emerging  fluxes calculated in a model with enhanced abundance of the selected
element by a factor of $10$ and the reference one. Silicon redistributes the far
UV flux to the spectral regions with $\lambda>1700\,\AA$. The influence of
chromium and iron on the emerging flux is more complex, but in general, these
elements redistribute the flux from the far UV region to the spectral region
with $\lambda\gtrsim2900\,\AA$. Magnesium and manganese do not affect the
emergent flux significantly.

\begin{figure}[tp]
\centering
\resizebox{\hsize}{!}{\includegraphics{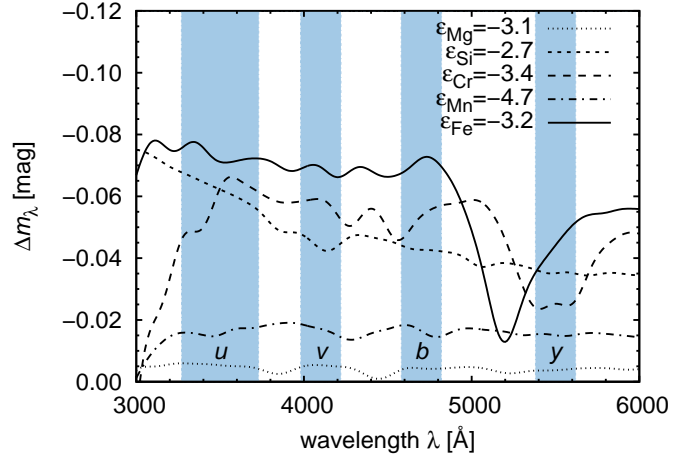}}
\caption{Magnitude difference $\Delta m_\lambda$ between the emerging fluxes
calculated with an enhanced abundance (by a factor of $10$)
of a selected element and the reference flux
$H_\lambda^\text{ref}$ (Eq.~\ref{tokmagroz}).
The fluxes were smoothed by a Gaussian filter with a dispersion
of $100\,$\AA.}
\label{magtoky}
\end{figure}

The flux changes caused by the heavier elements manifest themselves
in the changes of the apparent magnitude. This can be seen from the plot of the
magnitude difference
\begin{equation}
\label{tokmagroz}
\Delta m_\lambda=-2.5\log\zav{
\frac{H_\lambda(\varepsilon_\text{Mg},\varepsilon_\text{Si},
\varepsilon_\text{Cr},\varepsilon_\text{Mn},\varepsilon_\text{Fe})}
{H_\lambda^\text{ref}}},
\end{equation}
in Fig.~\ref{magtoky}. $H_\lambda^\text{ref}$ is the reference flux calculated
for $\varepsilon_\text{Mg}=-4.1$, $\varepsilon_\text{Si}=-3.7$,
$\varepsilon_\text{Cr}=-4.4$, $\varepsilon_\text{Mn}=-5.7$, and
$\varepsilon_\text{Fe}=-4.2$. For silicon the magnitude difference decreases
with increasing wavelength. There is a minimum of $\Delta m_\lambda$ at about
5200\,\AA\ for iron and at 5400--5600\,\AA\ for chromium, supporting the
conclusion that these elements significantly contribute to the flux depression
at these wavelengths \citep{kupzen,preslo,myhr7224}. As follows already from
Fig.~\ref{prvtoky}, magnesium and manganese do not significantly affect the
emerging fluxes in the case studied here.

\section{Predicted broad-band variations}

\begin{figure}[t]
\centering
\resizebox{\hsize}{!}{\includegraphics{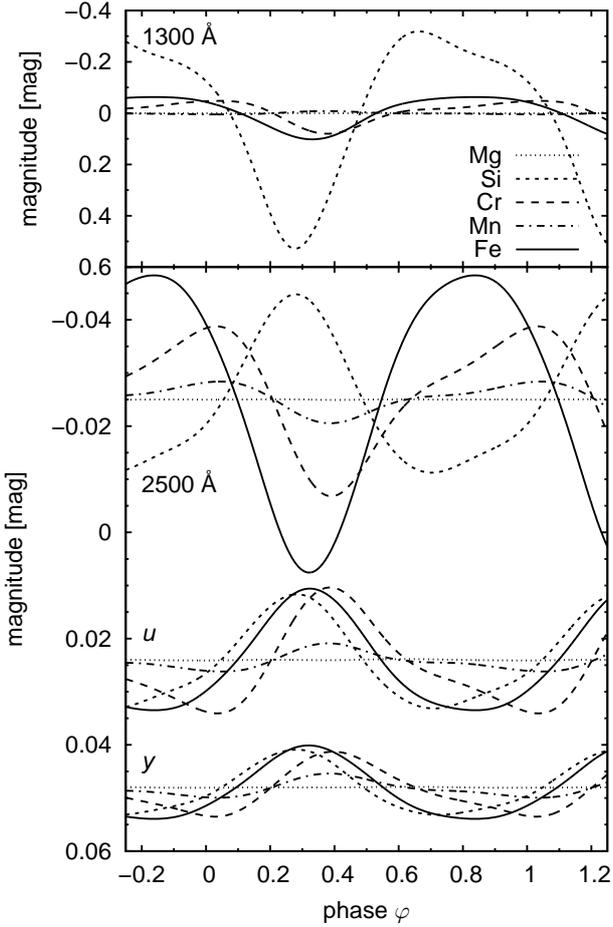}}
\caption{Predicted UV and visual flux variations of \hvezda\
calculated using abundance maps of one element alone. The abundance of
other elements was fixed. Light curves for selected wavelenghts in the UV
domain (denoted in the graph) were calculated using a Gaussian filter with
a dispersion of $100\,\AA$. The light variations were calculated for the
$u$ and $y$ filters of the Str\"omgren photometric system. Light curves in
individual bands were vertically shifted to better show the
variability.}
\label{prv_hvvel}
\end{figure}

\begin{figure}[t]
\centering
\resizebox{\hsize}{!}{\includegraphics{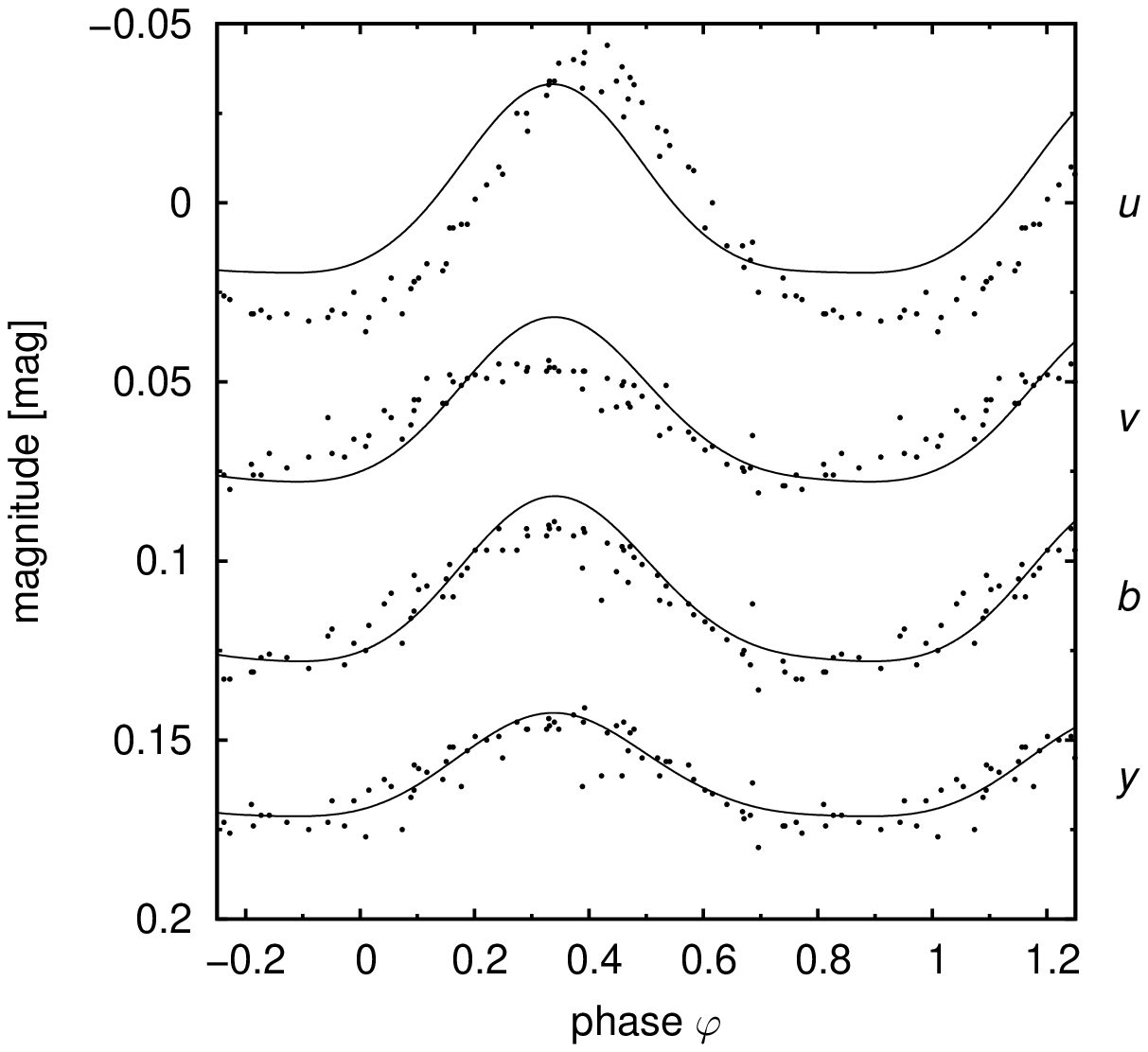}}
\caption{Predicted light variations of \hvezda\ (solid lines) in the Str\"omgren
photometric system computed taking the surface distribution of all elements.
The observed light variations (dots) are taken from \citet{aktax}. The light
curves in individual filters were vertically shifted to better show the
light variability.}
\label{cuvir_hvvel}
\end{figure}

Different parts of the stellar surface are exposed to the observer as a result
of the stellar rotation, which causes the stellar light variability. To account
for the stellar rotation, predicted light curves were calculated from
Eq.~\ref{velik} as a function of the rotational phase.

In Fig.~\ref{prv_hvvel}
we provide the light variations calculated from the abundance map of one element
alone assuming a fixed abundance of other elements
($\varepsilon_\text{Mg}=-4.1$, $\varepsilon_\text{Si}=-3.7$,
$\varepsilon_\text{Cr}=-5.9$, $\varepsilon_\text{Mn}=-5.7$, and
$\varepsilon_\text{Fe}=-4.2)$. From Fig.~\ref{prv_hvvel} it follows that mainly
silicon, chromium, iron, and partly manganese cause the light variability. These
elements are able to affect the emerging flux, and there is a large overabundance
of these elements in the spots. The light curves of individual elements reflect
the variations of line equivalent widths. As a result of the different surface
distribution of individual elements, the line equivalent widths curves and also
individual light curves in Fig.~\ref{prv_hvvel} differ in shape.

Silicon bound-free absorption dominates the flux distribution below roughly
$1700\,\AA$. The absorbed flux is redistributed to longer wavelengths,
consequently, the silicon flux variations at $1300\,\AA$ and $2500\,\AA$ are in
anti-phase in Fig.~\ref{prv_hvvel}. For similar reasons, the UV flux variations
and the visual light variations are in anthi-phase for chromium,
manganese, and iron.

\begin{figure*}[t]
\centering \resizebox{0.49\hsize}{!}{\includegraphics{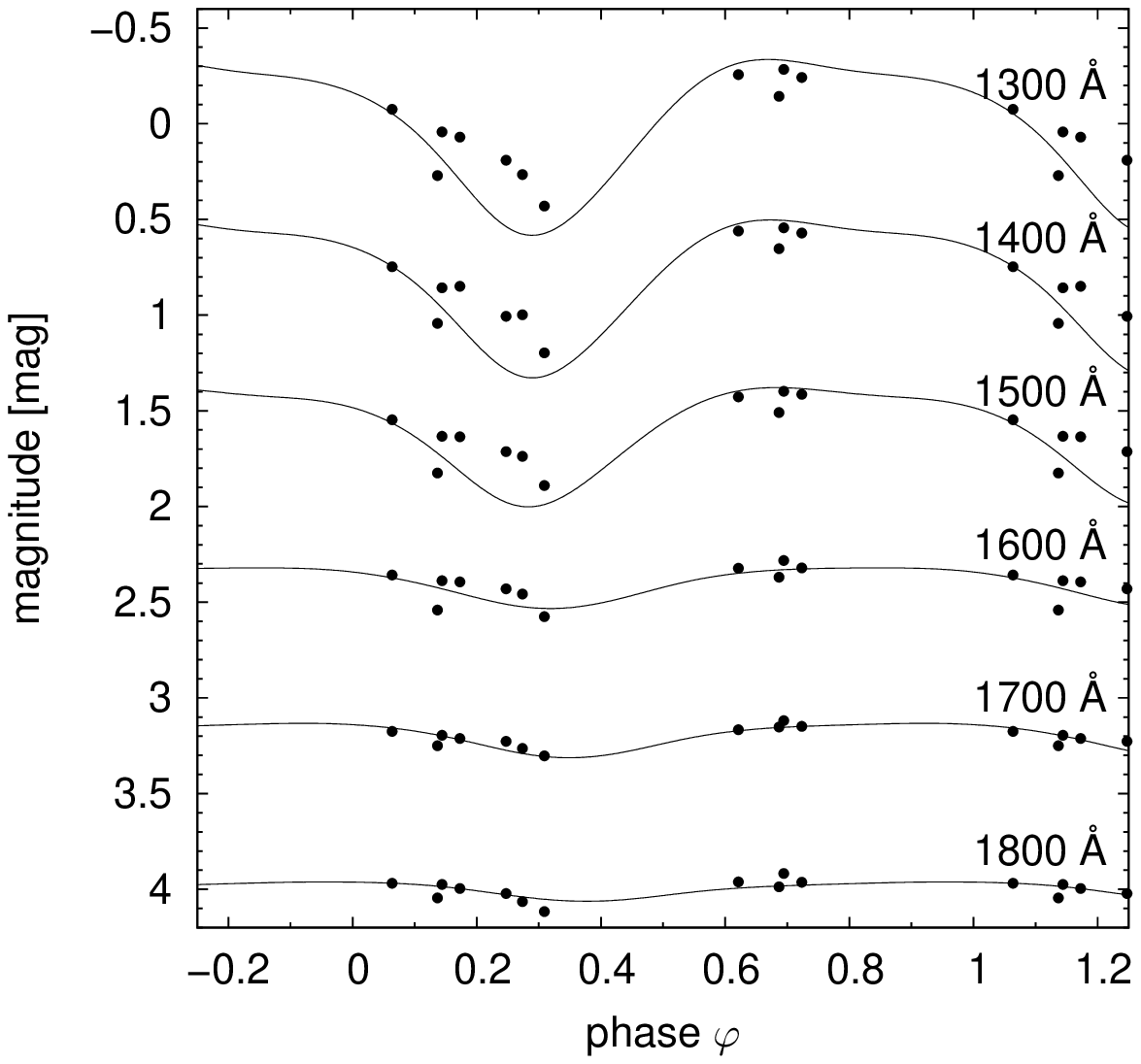}}
\centering \resizebox{0.49\hsize}{!}{\includegraphics{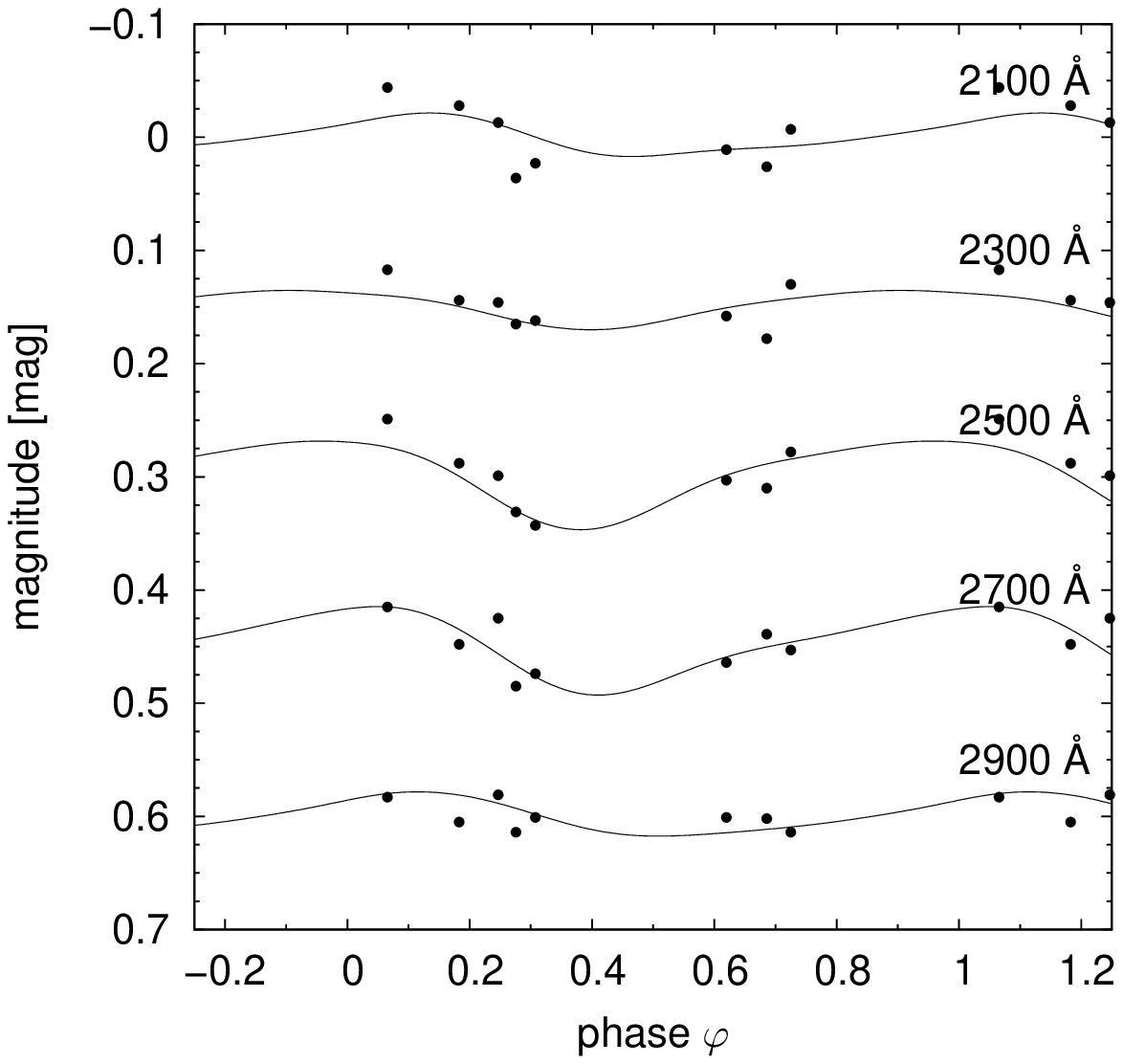}}
\caption{Comparison of the predicted (solid line) and observed (dots) UV
light variations for different wavelengths (marked in the graph). The
light curves were derived from IUE spectra using a Gaussian filter with a
dispersion of 25\,\AA\ (for SWP spectra) and 100\,\AA\ (LWR spectra).
Curves for individual wavelengths were vertically shifted to
better show the variability.
}
\label{fuvnuv}
\end{figure*}

Taking into account the surface distribution of all significant elements (see
Table~\ref{esit}), the observed and predicted light curves in the $b$ and $y$
filters of the Str\"omgren photometric system agree well in
Fig.~\ref{cuvir_hvvel}. The observed curves have slightly lower amplitudes than
the predicted curves in the $v$ filter, whereas there is also a slight phase
shift between these curves in the $u$ filter. A similar behaviour also appears
in the UV light curves in Fig.~\ref{fuvnuv}. Here we plot the observed and
predicted light curves, which were processed using a Gaussian filter. The
observed and predicted light curves agree very well in the long-wavelength part
of the UV spectrum with $\lambda\geq1600\,\AA$, whereas the observed curves have
lower amplitudes than the predicted curves in the far-UV domain with
$\lambda\leq1500\,\AA$.

\begin{figure}[t]
\centering
\resizebox{\hsize}{!}{\includegraphics{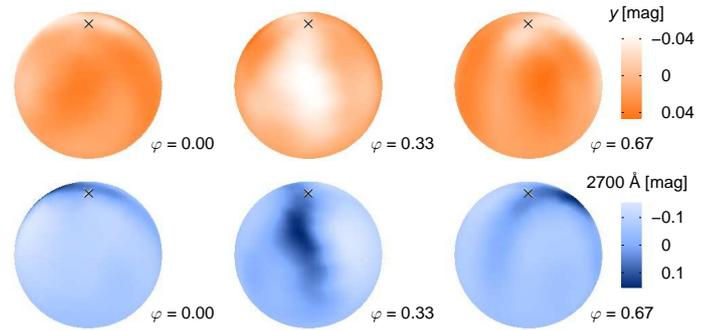}}
\caption{Emerging intensity from individual surface elements of \hvezda\ at
various rotational phases. Upper panel: visible $y$ band. Lower panel: UV band
centred at $2700\,$\AA. Both for $\mu=1$.}
\label{cuvir_povrch}
\end{figure}

The inhomogeneous surface distribution of individual elements causes bright
spots on the stellar surface. The spots, whose surface distribution can be
derived using abundance maps and model atmospheres (see
Fig.~\ref{cuvir_povrch}), cause the light variability. Given the relative
proximity of the star, \hvezda\ can be an ideal candidate for interferometric
analysis \citep[cf.][]{shuinfer}.

\section{Ultraviolet flux variations}
\label{uv}

We have shown that most of the light variability of \hvezda\ can be explained
by the flux redistribution from far UV to the near UV and visible regions.
This is supported by the comparison of UV and visual light curves, which are
typically in anti-phase. However, there is some disagreement between
observed and predicted light curves. The predicted $u$-amplitude is too low
(Fig.~\ref{cuvir_hvvel}) and, in addition, the predicted variations in the
far-UV are stronger than the observed variations (Fig.~\ref{fuvnuv}). Further
discussion of ultraviolet fluxes is necessary to understand the origin of these
differences.

\begin{figure*}[t]
\centering \resizebox{0.75\hsize}{!}{\includegraphics{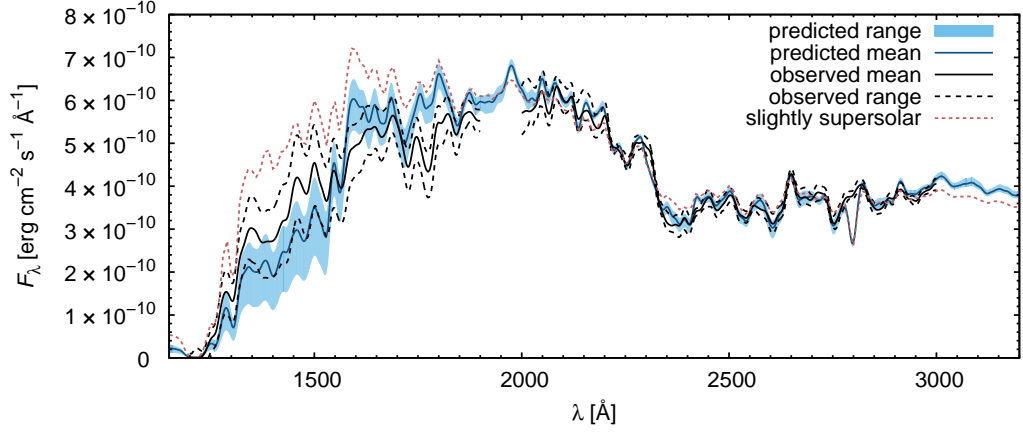}}
\caption{Comparison of the predicted flux (mean and its variation, blue)
with corresponding observed quantities (black). The predicted and observed
fluxes (IUE) were smoothed by a Gaussian filter with a dispersion of $10\,$\AA.
Overplotted is the flux calculated for a slightly supersolar chemical
composition with $\varepsilon_\text{Mg}=-4.1$, $\varepsilon_\text{Si}=-3.7$,
$\varepsilon_\text{Cr}=-5.9$, $\varepsilon_\text{Mn}=-5.7$, and
$\varepsilon_\text{Fe}=-4.2$.}
\label{ptok}
\end{figure*}

In Fig.~\ref{ptok} we compare the observed and the predicted fluxes smoothed by
a Gaussian filter with a dispersion of $10\,$\AA. We normalized the predicted
fluxes by a multiplicative factor, which yields the best match between
observations and predictions in Fig.~\ref{ptok}. The factor was kept fixed for
all wavelengths in all subsequent calculations.

From the UV flux normalization we derived the stellar radius
$R=4.34\pm0.05\,R_\odot$ assuming a distance of $50.8\pm0.4\,\text{pc}$
\citep{hipik}. This is significantly higher than the $2.2\,R_\odot$ reported
by \citet{pafra} or the $3.3\,R_\odot$ derived by \citet{babu}. On the other hand,
our value agrees with the radius of $4.3\,R_\odot$ adopted by \citet{ryze}.
Moreover, our derived radius yields with the rotational period and inclination
given in Table~\ref{hvezda} the projection of the rotational velocity
$v_\text{rot} \sin i = 53\,\text{km}\,\text{s}^{-1}$,
which agrees with observations (see Table~\ref{hvezda}).

The mean predicted and observed fluxes agree very well in the near-UV region
with $\lambda>2000\,\AA$ (see Fig.~\ref{ptok}). However, the mean predicted flux
is slightly lower than the observed one in the region $1250-1500\,\AA$. The
opposite is true in the region between $1700-1900\,\AA$. On the other hand, the
amplitude of the flux variations agrees well with observations in the whole
UV domain except for the region of $2000-2300\,\AA$.

\renewcommand\dbltopfraction{1.0}
\renewcommand\dblfloatpagefraction{1.}

\begin{figure*}[t]
\centering \resizebox{0.9\hsize}{!}{\includegraphics{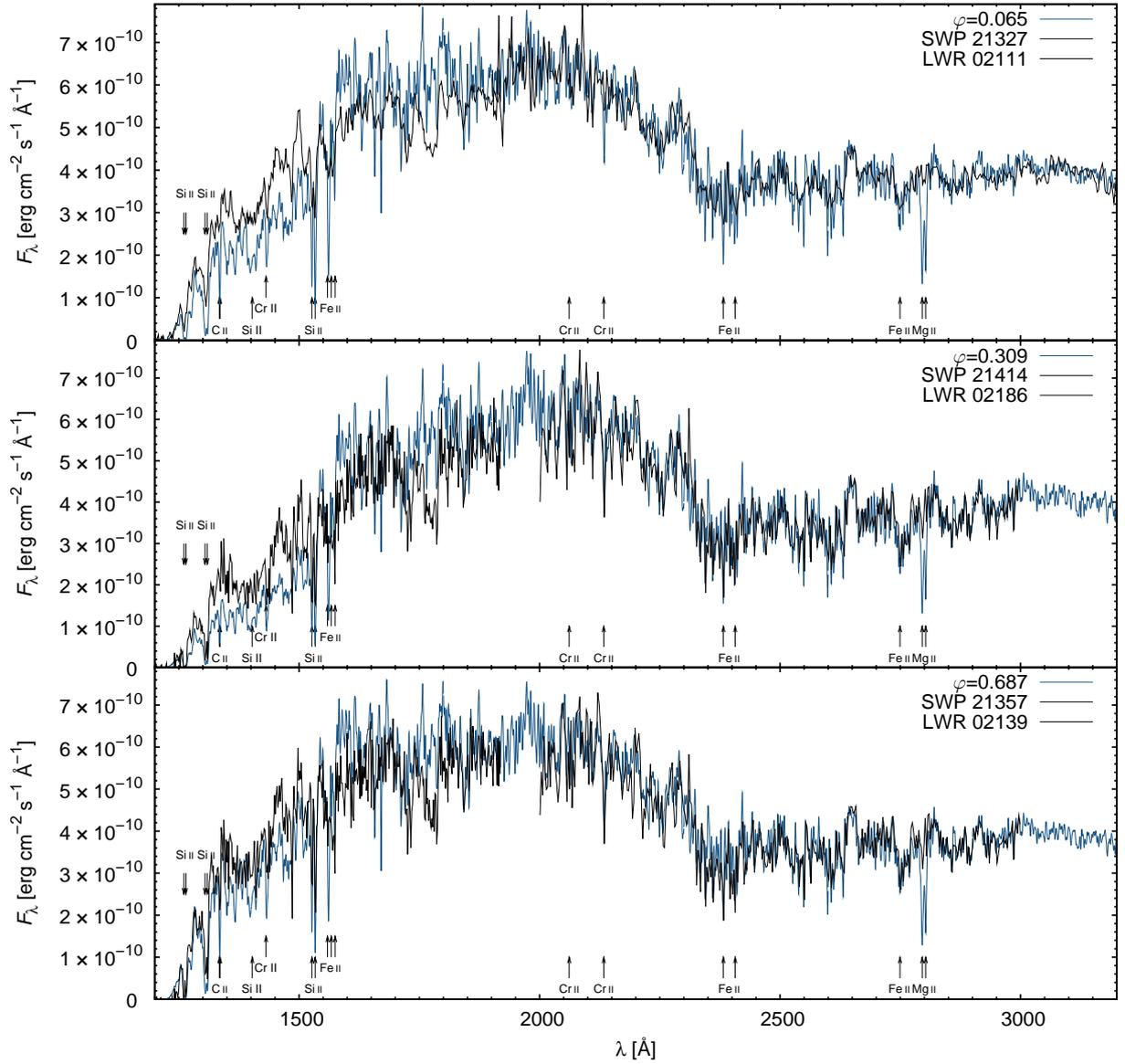}}
\caption{Predicted and observed (IUE) flux in selected phases. Predicted fluxes
were smoothed by a Gaussian filter with a dispersion of 1.3\,\AA. Individual
strong lines and iron line blends are identified.}
\label{mono}
\end{figure*}

This is further shown in Fig.~\ref{mono}, where we compare predicted and
observed fluxes for selected phases. The fluxes agree very well in the near-UV
region with $\lambda>1900\,\AA$. The shape of the pseudo-continuum (formed
by thousands of metallic lines) and the strongest line blends of mostly chromium
and iron agree very well. The only exception is the \ion{Mg}{ii} doublet at
$2800\,\AA$, whose line depth is overestimated by the spectrum synthesis.

The fluxes in the far-UV region agree less well. The predicted flux is
underestimated in the far UV region with $\lambda<1500\,\AA$ with an exception
of the phase around $\varphi=0.7$. Moreover, the observations show a stronger
line absorption in the region $1700-1800\,\AA$, which is not explained by the
models.

The fact that the observed and predicted fluxes most strongly disagree in the
region with $\lambda<1500\,\AA$, where the silicon strongly influences the flux
distribution (see Fig.~\ref{prvtoky}) points to a possible problem with the silicon
abundance map. This might be connected with the relatively few spectra
that were used to derive the maps (13). This might also explain the shift between
predicted and observed UV light curves in the bands with $\lambda\leq1500\,\AA$
(see Fig.~\ref{fuvnuv}). On the other hand, the stronger line absorption in the
region $1700-1800\,\AA$ is connected with missing line opacity in the models,
either caused by some element that was excluded from the mapping or by an incompleteness
of the current line lists.

A too strong predicted \ion{Mg}{ii} doublet at $2800\,\AA$ can be caused by a
vertical abundance distribution of magnesium. This is a resonance
doublet,
whereas the 4481\,\AA\ line used by \citet{kuschnigthesis} for surface
abundance mapping originates from excited levels, where the lower-level energy
of the corresponding transition is 8.9\,eV. A similar effect, but weaker, is also visible in some \ion{Si}{ii} lines. Vertical
abundance gradients are frequently found in other CP stars
\citep[e.g.,][]{lukory,rykochba,mahlad,balan}. Moreover, the carbon line blend
at 1335\,\AA\ shows an equivalent width variability, which suggests an inhomogeneous
surface distribution of this element.

\renewcommand\dbltopfraction{.9}
\renewcommand\dblfloatpagefraction{.95}

\begin{figure}[tp]
\centering \resizebox{\hsize}{!}{\includegraphics{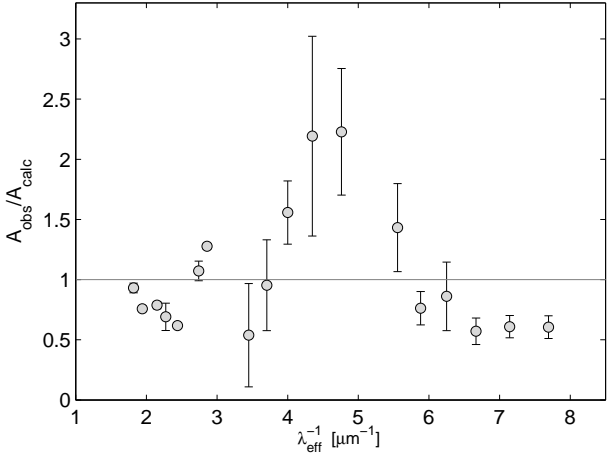}}
\caption{{Comparison of the amplitudes of the observed photometric variability and the variability predicted by the physical model. The real variability is in most cases weaker with some exception when it is strengthened by the role of the secondary spot at the phase 0.11.}}
\label{amplratio}
\end{figure}

From Fig.~\ref{ptok} it follows that even the mean observed UV flux does not
correspond to the solar chemical composition. Especially the observed far-UV
flux is significantly lower than the solar flux. This flux decline can presumably
not be modelled without reliably knowing  the surface abundance distribution. This
makes the temperature determination of CP stars from the UV spectra problematic.

\section{Detailed comparison of observed and predicted light curves}

We conclude that the predicted light curves explain the nature of the
photometric variability of \hvezda, but the correspondence is not
excellent. The physical model, for example, does not agree with all results obtained by
the phenomenological model described and discussed in Sect.\,\ref{fenomodel},
especially the incidence and properties of the second photometric spot centred
at the phase 0.11.

Although we see a large scatter in the ratio of the observed and predicted
amplitudes of light variations (see Fig.\,\ref{amplratio}), the real light
variability of the star is weaker than predicted. This might
be because some important sources of photometric variability
were neglected, because
including additional variability sources results in a less important role of the remaining sources. The other possibility is that
the variability caused by silicon is too large. Nevertheless, there are some
interesting exceptions -- the relatively strong enhancement of the observed
amplitude in the photometric band $u$, where some mechanisms connected with the
nearby Balmer jump might be important.

\begin{figure}[tp]
\centering \resizebox{\hsize}{!}{\includegraphics{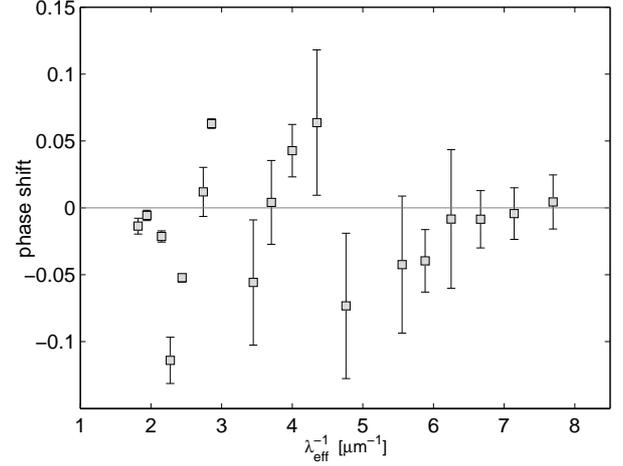}}
\caption{Dependence of the phase shift between the observed and predicted light
curves. Note the well documented jump in the shift in the region of the Balmer jump. See also the shapes of observed and predicted light curves in Fig.\,\ref{cuvir_hvvel}.}
\label{shift}
\end{figure}

The phase shift between the observed and predicted light curves in
Fig.\,\ref{shift} provides additional information about the agreement between
observed and predicted light curves. The shift in the optical region decreases
with increasing frequency, and in the region of the Balmer jump, it suddenly
switches to a large positive shift. The phenomenological model explains this
phenomenon by strong changes in the amplitude of the secondary spot, which
reaches its highest brightness before the Balmer jump and almost
vanishes after the jump. Trends in the shift in other parts of the figure can be
also explained by the wavelength dependency of the contrast of the secondary
spot.

The predicted light curves are based on the horizontal elemental abundance
variations derived from observations. However, CP stars also
show vertical
abundance gradients, which in \hvezda\ are probably evidenced in some UV line
profiles. It is possible that the emerging flux at different wavelengths
originates in different vertical regions. Consequently, the vertical abundance
gradients in the atmosphere may in principle affect the light curves, and neglecting vertical abundance gradients may be one of the reasons for
the disagreement between theoretical and observed light curves.

\section{Conclusions}

We successfully simulated the UV and visual flux variability of the chemically
peculiar star \hvezda. We assumed that the variability is caused by the
inhomogeneous surface distribution of chemical elements. We used model
atmospheres calculated with appropriate abundances corresponding to the Doppler
maps of \citet{kuschnigthesis} to predict the light variability.

We also used our phenomenological model to improve the rotational period of the
star and to model the light curves in the spectral region 1300 -- 5500\,\AA. The
phenomenological model assumes two photometric spots with distinct optical
characteristics. We determined the geometry of the magnetic field and showed
that there is no obvious connection between the field and photometric spots on
the stellar surface.

The numerical simulations based on the Doppler maps and model atmospheres
explain most of the variability of \hvezda. Heavier elements influence the
emerging flux through flux redistribution mainly from the far-UV to near-UV and
visible regions. The bound-free transitions of silicon and the bound-bound
transitions of chromium, iron, and to a lesser extent manganese mostly cause the
flux redistribution and variability in \hvezda. As a result of the flux
redistribution and inhomogeneous elemental surface distribution, the individual
surface elements display different brightnesses in individual spectral regions.
The total observed flux, which is given by integrating over the whole
stellar surface, is modulated by the stellar rotation. Resulting simulated light
curves were compared with synthetic colours derived from IUE spectroscopy and
with observed variability in bands of the Str\"omgren photometric system.

The inhomogeneous surface distribution of silicon, chromium, manganese, and iron
is able to explain most of the observed UV and visible flux variations. We
reproduced the anti-phase behaviour of the light curves in the far-UV and visible
regions. The light curve amplitude is about a few hundredths of a
magnitude in the visual domain, while it reaches nearly 1\,mag in the UV. There
remains some minor disagreement between the predicted and observed light curves,
particularly in the $u$ colour of the Str\"omgren photometric system or in the
far-UV band at $1300\,\AA$.

Our models reproduce the observed spectral energy distribution in the near-UV
region in detail. The absorption features that resemble strong lines in this
region are in fact mostly blends of numerous iron and chromium lines. The
spectral energy distribution in the far-UV region is only well reproduced around
the phase of $\varphi=0.7$. There are minor differences in other phases, which
may be connected with a coarser phase coverage of the spectra used to prepare
the silicon abundance maps. Other discrepancies in the far-UV may be attributed
to some missing line opacity in the models.

The UV spectral analysis is very well suited for detecting vertical
abundance gradients. The UV region comprises many strong lines that originate from
states with low excitation potential, which provides information about the upper parts of
the atmosphere. In combination with an optical analysis (on which the Doppler
maps are based), which also yields information from the lower atmospheric
parts, we were able to detect the vertical abundance gradient of magnesium and
possibly silicon.

Our study provides additional evidence that the light variability of
chemically peculiar stars is in most cases caused by the inhomogeneous surface
distribution of individual chemical elements and flux redistribution and is modulated
by the stellar rotation. For cooler chemically peculiar stars in addition to silicon
and iron, other iron-peak elements also influence the light variability, in our
case, mostly chromium. The comparison of the observed and predicted spectral
energy distribution was able to reveal all missing opacities in the model
atmospheres.

\begin{acknowledgements}
This research was partly based on the IUE data derived from the INES database
using the SPLAT package and has made use of the NIST spectral line database
\citep{nist}. This work was supported by the grant GA \v{C}R P209/12/0217 and by
a mobility grant of M\v SMT 7AMB14AT015 (WTZ CZ 09/2014). T.~L\"uftinger
acknowledges the support by the FWF NFN project S116601-N16 and the related FWF
NFN subproject, S116 604-N16.
\end{acknowledgements}

\end{document}